\documentclass[a4paper,10pt]{article}
\usepackage{amssymb,amsmath}

\newcommand{\cyl}[1]{$\Delta^*_{\varepsilon (#1)}$}
\begin{document}

\begin{flushright}
  QMUL-PH-07-24\\
  DESY-07-188
\end{flushright}

\begin{center}
  \Large{\bf BCFT and Ribbon Graphs as tools for open/closed string dualities}
\end{center}

\begin{center}
  \underline{Valeria L. Gili}$^{a,}$\footnote{Contributing author. Email: V.Gili@qmul.ac.uk}, 
Mauro Carfora$^{b,}$\footnote{Email: Mauro.Carfora@pv.infn.it}, 
Claudio Dappiaggi$^{c,}$\footnote{Email: Claudio.Dappiaggi@desy.de}

\vspace{1cm}

{\footnotesize
\noindent$^{a}$~Centre for Research in String Theory,
Department of Physics,\\
Queen Mary, University of London,\\
Mile End Road, London, 
E1 4NS, UK

\vspace{0.2cm}
\noindent$^{b}$~Dipartimento di Fisica Nucleare e Teorica,\\
Universit\`a degli Studi di Pavia and INFN,\\
Via A.Bassi, 6,
I-27100, Pavia, Italy

\vspace{0.2cm}
\noindent$^{c}$~II. Institut f\"ur Theoretische Physik,\\
Universit\"{a}t Hamburg,\\[0pt]
\vspace{-0.1cm}
Luruper Chaussee 149,
22761 Hamburg, Deutschland}

\vspace{1cm}
\emph{\small To appear in the Proceeding of the 7th International Workshop\\
"Lie Theory and Its Applications in Physics" (LT-7),\\
18-24 June 2007, Varna, Bulgaria. }
\vspace{1cm}

\begin{abstract}
In the framework of simplicial models, we construct and we fully characterize a
scalar boundary conformal field theory on a triangulated Riemann surface. The
results are analysed from a string theory perspective as tools to deal with
open/closed string
dualities.
\end{abstract}
\end{center}
\vspace{1cm}


Simplicial techniques have being subject of renewed attention since it
was clarified that they can play a role in the worldsheet description of
open/closed string dualities \cite{SimplString}. In this connection,
the hypothesis that the link between SYM gauge and closed string
worldsheet dynamics could be explained through the combinatorial data
associated to Strebel differentials, suggests that we are
undercovering some deep discrete foundation of string dualities, which
however is still far from being understood.

Aiming to give further insights in this topic, we decided to explore
the connection between combinatorial data and string dualities from a
more general standpoint. We considered a geometrical set-up in which
ribbon graphs are dual to triangulations with localized curvature
defects.  These provide a natural order parameter which allows to map
a $N$-punctured closed Riemann surface into an open one with $N$
boundary components.  An example of such a mapping has been given, in
an hyperbolic setting, in \cite{Carfora:2006nj}. A different
construction was obtained describing a random Regge triangulation
(RRT) \cite{Carfora:2002rn} as the uniformization of an open Riemann
surface $M_\partial$ with a set of $N$ annuli, \cyl{p}, $p = 1,
\ldots, N$ each of which is defined in the neighborhood of the $p$-th
vertex of the original triangulation. Each annulus is endowed with a
correspondent Euclidean cylindrical metric and, via a conformal
mapping can be equivalently interpreted as a cylinder of \emph{finite
  height}.  The decorated Riemann surface is subsequently constructed
glueing the above local uniformizations along the pattern defined by
the ribbon graph baricentrically dual to the parent triangulation.

This geometrical setup, which trades the localised curvature degrees
of freedom of the parent triangulation into modular data of the new
discrete surface, is simpler than that analysed in
\cite{Carfora:2006nj}, but it can be dynamically coupled with matter
field theory. In \cite{Carfora:2007tk} we showed that this leads to
the definition of a new kinematical background in which it could be
possible to investigate dynamical processes typical of open/closed
string dualities.

\section{Boundary Insertion Operators}

The geometry we dealt with is characterized by $N$ cylinders of finite
heights, which can be interpreted as open string worldsheet, connected
through their inner boundary to a ribbon graph. Hence, the latter is
the natural \emph{locus} where $N$ copies of a given Boundary
Conformal Field Theory, each defined on a single cylindrical end,
interact.

The quantization of a BCFT on a cylindrical domain is a delicate issue
relying on the knowledge of the data associated to the correspondent
\emph{bulk} CFT.  This is uniquely characterized by two copies of a
chiral algebra, $\mathcal{W}$ and $\overline{\mathcal{W}}$. Their
generators, respectively $W^a_n$ and $\overline{W}^a_n$, are the
Laurent's modes of the holomorphic and antiholomorphic chiral fields
of the theory, which we will call $W^a(\zeta)$ and
$\overline{W}^a(\overline{\zeta})$ respectively.  The extension of
such a bulk CFT to a BCFT on \cyl{p} consists in the choice of a
boundary condition $A(p)$ on $\partial\mbox{\cyl{p}}$. This is usually
done by specifying a glueing automorphism, $\Omega_{A(p)}$, which,
relating the holomorphic and anti-holomorphic chiral fields on the
boundary itself, avoids flux of informations through it. This process
allows to define, out of $W^a(\zeta)$ and
$\overline{W}^a(\overline{\zeta})$, a single chiral field,
$\mathbb{W}_{\Omega(p)}(\zeta)$. This is continuous on the full
complex plane and it contains the information about the boundary
condition.  Its Laurent modes close a single copy of the chiral
algebra, whose irreducible representations define the Hilbert space of
states of the boundary theory.

In the framework dual to a RRT, each $(p,q)$-edge of the ribbon graph
is connected to two cylindrical ends, \cyl{p} and \cyl{q}. Hence, its
oriented boundaries are decorated by two different boundary
conditions, say $A(p)$ and $B(q)$ respectively. In this
picture, we do not have a jump between two boundary conditions taking
place in a precise boundary's point. Therefore, we cannot apply standard
BCFT rules which, assuming the existence of a vacuum state not
invariant under translations, allow a boundary condition to change
along a single boundary component.

In order to mediate between \emph{adjacent} boundary conditions, we
have thus introduced a different process which, restoring the flux of
information trough the boundaries connected to the ribbon graph,
allows to describe the interaction of the different BCFTs on the
ribbon graph. In the limit when the thickness of the graph goes to
zero, we introduced, for each pair $(p,q)$ of adjacent BCFTs, a
further glueing automorphism $\Omega_{A(p) B(q)}$. This deforms
$\mathbb{W}_{\Omega_{A(p)}}$ into $\mathbb{W}_{\Omega{B(q)}}$
continuously on the graph's edge and viceversa.  This ultimately means
that we are associating to each $(p,q)$ edge of the ribbon graph a
unique copy of the chiral algebra. Since the latter encodes the
information about both the boundary conditions applied on the adjacent
polytopes, it is natural to choose the highest weight operators of its
irreducible representation as objects which mediates between adjacent
boundary conditions. We called this new class of fields Boundary
Insertion Operators.

\section{BIO in the rational limit}
BIOs as described in last section are purely formal objects. However
we characterized explicitly their algebraic and analytic structure in
particular limits of the BCFT.  We considered a $D$-dimensional
bosonic field theory, $X^\alpha: M_\partial \rightarrow \mathcal{T}$,
$\alpha = 1, \ldots, D$, where the geometry of the target space
$\mathcal{T}$ is encoded, in the $k$-th cylindrical end, by the
background matrix $E(k) = G(k) + B(k)$. We focused on a flat toroidal
background, where the $D$ directions are compactified and $E(k)$'s
components are $X$-independent. Within this framework, each
cylindrical end is an open-string worldsheet whose outer boundary can
lay on a stack of $N_c$ D-branes. While, on the one hand, the latters
allow to decorate each open string with a suitable assignment of
Chan-Paton factors, on the other hand they act as source of gauge
fields as well.  In the case of static brane and constant gauge field
strength, the gauge dynamic can be encoded by the toroidal background
via the identification $F_{\alpha\beta} = \frac{1}{4\pi}
B_{\alpha\beta}$.  This translates the problem of coupling the model
with a dynamic gauge field into picking up a special point in the
toroidal compactifications moduli space such that only the component
of the Kalb-Ramond field along the directions parallel to the brane
are non zero.

Moreover, if we break $U(N_c) \rightarrow U(1)^{N_c}$, and we fix the
metric and the antisymmetric field in function of the Cartan matrix of
a semi-simple simply laced Lie algebra of total rank $D$,
$\mathfrak{g}_D$, we are choosing those points in the moduli space
which are fixed under the action of the generalized $T$-duality group.
The resultant theory of $D$ compactified bosons turns out to be
rational, since it is quantum equivalent to the
$\hat{\mathfrak{g}}_{k=1}$-WZW model, where $\hat{\mathfrak{g}}_{k=1}$
is the affine extension of $\mathfrak{g}_D$ at level $k=1$.

In this connection, we showed that we can parameterize the entire set
of boundary conditions we can apply on the boundary of \cyl{k} by the
pair $[\Vert\hat{\omega}_I\rangle\rangle, \Gamma(p)]$. The first
element of the pair, $\Vert\hat{\omega}_I\rangle\rangle$, is a Cardy
boundary state of the WZW model. It is uniquely associated to the
$\hat{I}$-th level one irreps. of $\hat{\mathfrak{g}}_{k=1}$. The
second element, $\Gamma(p)$, is an element of the quotient of the
universal covering group of $\mathfrak{g}_D$ by its centre, {\small
  $\Gamma(p) \in \frac{G_D}{B(G_D)}$}.  This parameterization allowed
to give a description of the model as a deformation of the
$\hat{\mathfrak{g}}_{k=1}$-WZW model described \emph{\`a la Cardy} by
means of a precise boundary action {\small $S_\Gamma = \int du
  \Gamma_a J^a(u)$}.  In the last formula, coefficients $\Gamma_a$ are
defined via a suitable immersion of $\Gamma(p)$ into the universal
covering group $G_D$.
This description led to the characterization of the glueing
automorphism associated to the $(p,q)$-edge of the ribbon graph,
$\Omega_{[\hat{J}_2 \Gamma_2](q) [\hat{J}_1 \Gamma_1](p)}$, by means
of the fusion coefficient of the WZW model, {\small
  $\mathcal{N}_{\hat{J}_1(p) \hat{I}(p,q)}^{\hat{J}_2(q)}$}, and by a
deformation induced by the rotation $\Gamma(p,q)= \Gamma_2(q)
\Gamma_1(p)^{-1}$. As a consequence, we wrote the components of
Boundary Insertion Operators in the rational limit of the conformal
theory as {\small$
  \psi^{[\hat{J}_2,\,\Gamma_2](q)\,[\hat{J}_1,\,\Gamma_1](p)}
  _{[\hat{I},\,m](p,q)} \,=\, \sum_{n=0}^{\text{dim}|\hat{I}|}
  R^{\hat{I}(p,q)}_{m\,n(p,q)}(\Gamma_2{\Gamma_1}^{-1}) \,
  \psi_{[\hat{I},\,n](p,q)}^{\hat{J}_2(q)\,\hat{J}_1(p)}$}, where
{\small $\psi_{\hat{J}(p,q)}^{\hat{J}(p)\,\hat{J}(q)} (x(p,q)) \,=\,
  \mathcal{N}_{\hat{J}(p)\,\hat{I}(p,q)}^{\hat{J}(q)}
  \,\psi_{\hat{I}(p,q)}(x(p,q))$} and where {\small
  $R^{\hat{I}(p,q)}_{m\,n(p,q)}$} is the action of
$\Gamma_2{\Gamma_1}^{-1}$ in a given representation.  In this case,
BIOs algebra is set in terms of the unperturbed WZW model fusion
rules. This property is particularly manifest in the algebraic form for
the expansion coefficients of the OPE between {\small
  $\psi^{[\hat{J}_1,\,\Gamma_1](p)\,
    [\hat{J}_3,\,\Gamma_3](r)}_{\hat{I}_1(r,p)}$} and {\small
  $\psi^{[\hat{J}_3,\,\Gamma_3](r)\,
    [\hat{J}_2,\,\Gamma_2](q)}_{\hat{I}_2(q,r)}$} in terms of {\small
  $\psi^{[\hat{J}_1,\,\Gamma_1](p)\,
    [\hat{J}_2,\,\Gamma_2](q)}_{\hat{I}_3(q,p)}$}, namely {\small
  $\mathcal{C}_{\hat{I}_1(r,p)\,\hat{I}_2(q,r)\,\hat{I}_3(q,p)}^{\hat{J}_1(p)\,\hat{J}_3(r)\,\hat{J}_2(q)}$},
whose labellings are left untouched by the deformations.  The variable
connectivity of the ribbon graph, which has been inherited by the original
triangulation, allowed to identify the above coefficients with the
$\hat{\mathfrak{g}}$-WZW model fusion matrices
\cite{Runkel:1998pm,Behrend:1999bn,Felder:1999ka}, namely the quantum
6-$j$ symbols {\small $\mathcal{C}^{\hat{J}_1(p)\,\hat{J}_2(q)\,\hat{J}_3(s)}
  _{\hat{I}_1(q,p)\,\hat{I}_2(s,q)\,\hat{I}_3(s,p)}=\left\{
\begin{smallmatrix}
    \hat{I}_1(q,p) & \hat{J}_1(p) & \hat{J}_2(q)  \\
    \hat{J}_3(s) & \hat{I}_2(s,q) & \hat{I}_3(s,p)
  \end{smallmatrix}\right\}_{Q\,=\,e^{\frac{2 \pi i}{1 + h^\vee}}}$}\cite{AlvarezGaume:1988vr}.

\section{Discussion and Conclusions}
The last formula provides an explicit expression for the formal rules
describing the interplay among BCFT defined on the adjacent cylinders
which build the discrete surface $M_\partial$.  In this sense, it
completes the description of the dynamical coupling between the
discrete geometric set-up developed in \cite{Carfora:2002rn} and a
matter field theory. Out of this, in \cite{Carfora:2007tk} we wrote a
full amplitude on a fixed geometry specified by a choice for the
ribbon graph and for a set of localized curvature assignations.

Our aim for the future is to generalize this scenario to the
non-Abelian case, in which the $U(N_c)$ symmetry carried by the
D-Branes is not broken.  In this connection, the open string
interpretation would provide naturally a non-trivial gauge colouring
of the ribbon graph. This ultimately leads to the definition of a
genuine 't Hooft diagram and, hence, of new kinematical background in
which it would be possible to investigate dynamical processes
proper of open/closed string dualities.

\begin{center}
\subsection*{Acknowledgments}
\end{center}
  V.L.G. would like to thank the Foundation Boncompagni-Ludovisi,
  n\'ee bildt, for financially supporting her stay at Queen Mary,
  University of London. 

  The work of C. D. has been supported by the Alexander Von Humboldt
  Foundation.

\vspace{1cm}


\end{document}